\documentclass[a4paper,usenatbib]{mnras}
\usepackage{natbib}
\usepackage{graphicx, amssymb, amsmath}
\usepackage{comment}
\usepackage{xspace}
\usepackage{defs}
\usepackage{cleveref}
\usepackage{hyperref}
\usepackage{listings}
\usepackage{tabularx, booktabs}
\usepackage[T1]{fontenc}
\usepackage{ae,aecompl}
\usepackage{longtable}
\crefname{section}{§}{§§}
\newcommand{\Dn}{$D_{n}(4000)$\xspace}

\newcommand{\OII}{[\ion{O}{II}]\xspace}
\newcommand{\ewHd}{EW($H\delta$)\xspace}
\newcommand{\ewOII}{EW([\ion{O}{II}])\xspace}
\newcommand{\DoNothingA}[1]{#1}
\newcommand{\tbf}{\DoNothingA}

\title[Suppressed Star Formation in RX J0910]{Suppressed Star Formation by a Merging Cluster System} 
\author[A. S. Mansheim et al.]{
A. S. Mansheim,$^{1}$\thanks{E-mail: bclemaux@ucdavis.edu},
B. C. Lemaux$^{1}$,
A. R. Tomczak$^{1}$,
L. M. Lubin$^{1}$, 
N. Rumbaugh$^{2}$,\newauthor
P.-F. Wu$^{3}$,
R. R. Gal$^{4}$,
L. Shen$^{1}$,
W. A. Dawson$^{5}$,
G. K. Squires$^{6}$ \\
$^{1}$Physics Department, University of California, Davis, One Shields Avenue, Davis, CA 95616, USA\\
$^{4}$Institute for Astronomy, University of Hawai'i, 2680 Woodlawn Drive, HI 96822, USA\\
$^{3}$Max-Planck Institut f\"{u}r Astronomie, K\"{o}nigstuhl 17, D-69117, Heidelberg, Germany\\
$^{2}$National Center for Supercomputing Applications, University of Illinois, 1205 West Clark St., Urbana, IL, 61801, USA\\
$^{5}$Lawrence Livermore National Laboratory, 700 East Avenue, Livermore, CA, 94550, USA\\
$^{6}$California Institute of Technology / IPAC, M/S 314-6, 1200 E. California Blvd., Pasadena, CA, 91125, USA\\
}

\date{Accepted March 16th, 2017. Received March 16th, 2017; in original form September 14th, 2016}

\pubyear{2017}

\begin{document}
\label{firstpage}
\pagerange{\pageref{firstpage}--\pageref{lastpage}}
\maketitle

\begin{abstract}

We examine the effects of an impending cluster merger on galaxies in the large scale structure (LSS) RX J0910 at $z$ $=$1.105.
Using multi-wavelength data, including 102 spectral members drawn from the Observations of Redshift Evolution in Large Scale Environments (ORELSE) 
survey and precise photometric redshifts, 
we calculate star formation rates and map the specific star formation rate density of the LSS galaxies. 
These analyses along with an investigation of the color-magnitude properties of LSS galaxies indicate lower levels of star formation activity in the region between the merging clusters relative to the outskirts of the system.  
We suggest that gravitational tidal forces due to the potential of the merging halos \tbf{may be} the physical mechanism responsible for the observed suppression of star formation in galaxies caught between the merging clusters.

\end{abstract}

\begin{keywords}
galaxies: clusters: general -- galaxies: evolution -- galaxies: clusters: individual (J0910+5422) -- galaxies: clusters: individual: (J0910+5419)
\end{keywords}

\section{Introduction}\label{s:intro}

The environment in which a galaxy lives can strongly influence the pace and course of its evolution. 
By $z\sim1$ gravitational effects\tbf{, driven largely by the presence of dark matter,} have given rise to a spectrum of galaxy environments ranging in size from isolated galaxies to filaments and clusters. The dark matter halo exerts a differential gravitational (i.e., tidal) force over the length of a galaxy that can alter its internal processes and strip it of fuel for star formation. On a larger scale, gravity causes structures to merge and form increasingly complex large scale structures (LSSs). 
Cluster mergers, in particular, are the most cataclysmic manifestation of hierarchical structure formation, releasing energy on scales second only to the big bang. Merging clusters play an increasing role at higher redshifts, and subsequently lead in part to the properties observed in clusters today \citep[e.g.,][]{CohnWhite05}	

Though numerous studies on the effects of mergers on galaxy evolution have been performed, the results are conflicting. It has been indicated that slow (e.g., galaxy harassment or strangulation) and fast (e.g., ram pressure stripping \tbf{in the core)} processes associated with the merging event could either trigger \citep[e.g.,][]{MillerOwen2003}, 
quench \citep[e.g.,][]{P2004}
or have no effect on star formation \citep{Chung2009}, though with a small number of systems and a variety of methods employed for analyzing such effects. One of the greatest obstacles to studying these effects is that, in the diaspora after a first pass-through, all history of the clusters' initial states is erased. \citet{ali17a} found that, even with an exquisite, multiwavelegth data set and a dynamical simulation to constrain an ideal merger timeline, having no knowledge of the prior states of the galaxies proved insurmountable to connecting star formation to the merger event. 

In this study we investigate the pre-merging RX J0910 LSS at $z\sim1.10$ and perform a direct comparison of regions more and less affected by the impending merger. In RX J0910 we have the opportunity to analyze in detail a system that has not been fully disrupted by an initial pass-through, which minimizes the difficulties of membership contamination. 
Additionally, by conducting internal comparisons within the LSS and by comparing results from a variety of different metrics, we further minimize uncertainties associated with differing observational conditions, redshift, and the \tbf{large variation of} star formation rates observed at z$\sim$1 \citep[e.g.,][]{2007A&A...472..403T}. 


\begin{figure*}
          \begin{center}
          \includegraphics[height=70mm]{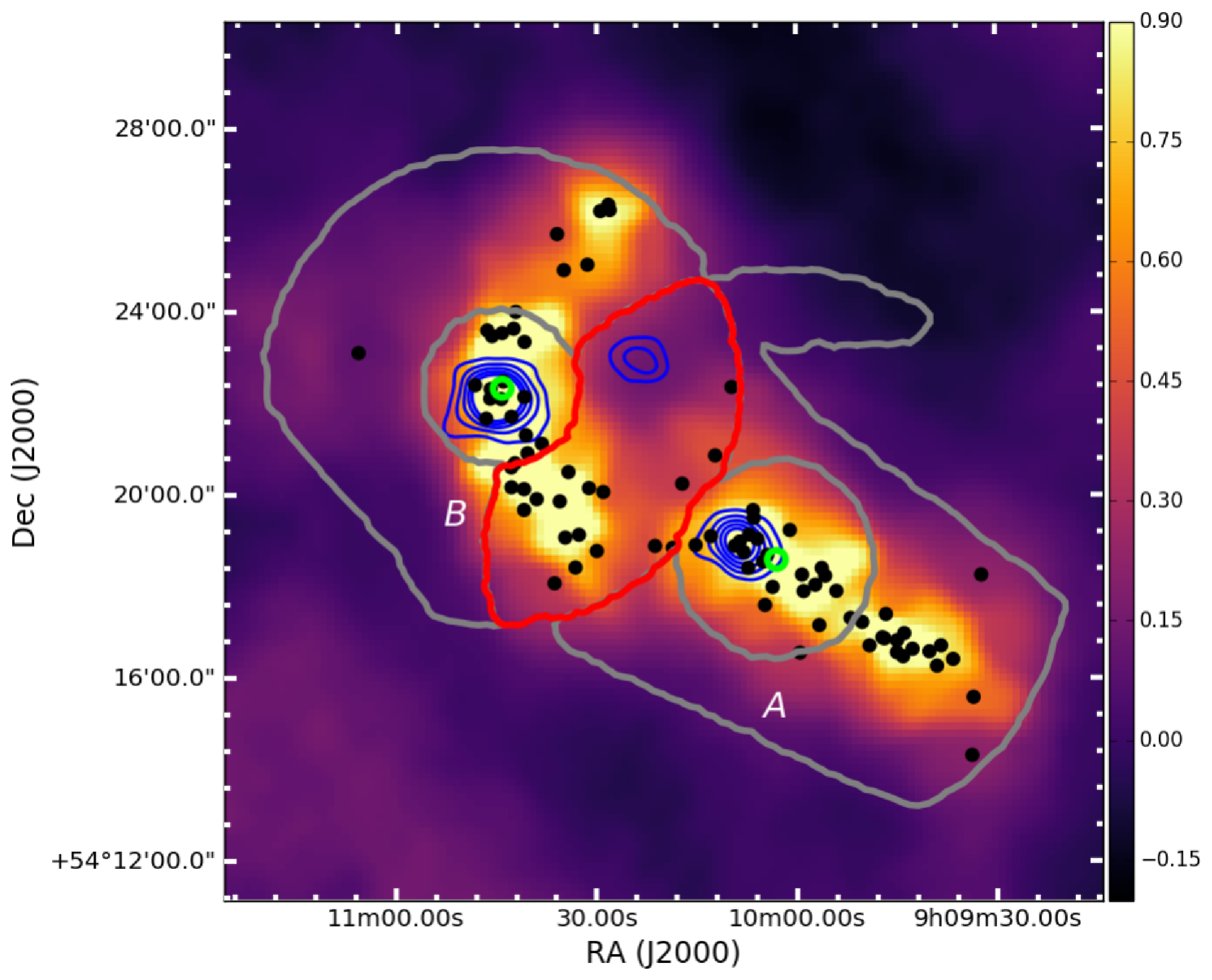}
          \includegraphics[height=70mm]{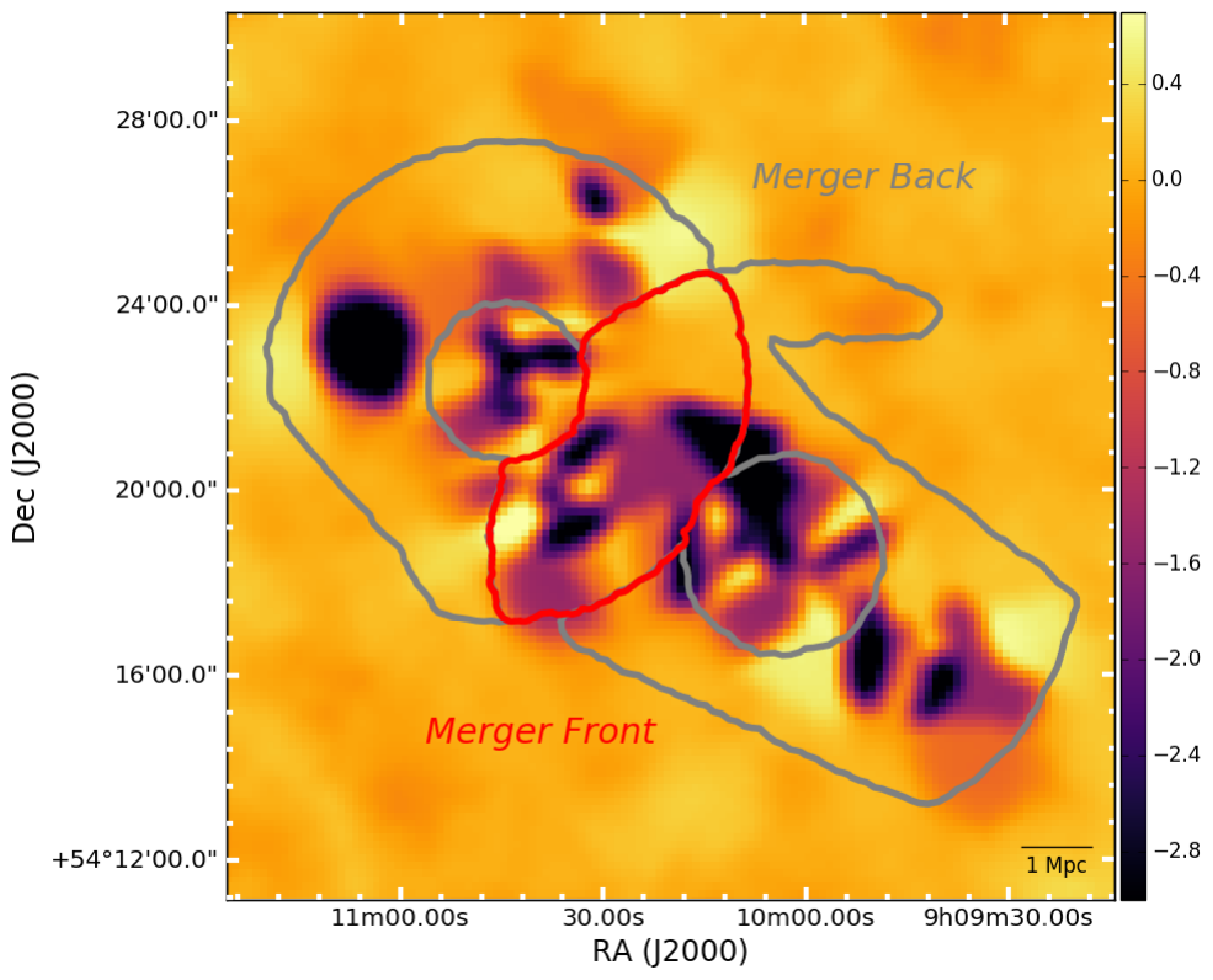}
          \end{center}
        \caption{
        \textit{Left:} Voronoi Monte Carlo map of galaxy overdensity in units of log(1+$\delta_{gal}$) for a $19.2'\times19.2'$ ($\sim$12.6 Mpc x 12.6 Mpc) field of view surrounding the two merging clusters, labeled A \tbf{(right, RXJ0910+5419, R$_{vir}$=1.01 $h^{-1}_{70}$ Mpc)} and \tbf{B (left, RXJ0910+5422, R$_{vir}$=0.82 $h^{-1}_{70}$ Mpc)}. Both panels are smoothed with a boxcar filter of $\sim$25$\arcsec$ ($\sim200$ kpc). Overlaid are the MB (gray) and MF \tbf{(red)} region boundaries defined in detail in \S \ref{s:results}. These regions exclude the area within 1.5$R_{vir}$ from the luminosity-weighted centers of clusters A and B, indicated by the overlaid small green circles, and extend to a maximum of 4$R_{vir}$ from these centers.
        \tbf{Black dots represent spectroscopically confirmed members. A $\beta$-profile smoothed \emph{Chandra} image made following \citet{2013ApJ...763..124R} was used to generate the blue contours which correspond to 2, 4, 6, 8 and 10$\times$ the image RMS.}
        \textit{Right:} VMC map representing SSFR$_{SED}$ overdensity per galaxy in units of log(1$+\delta_{ssfr/gal}$).
        \tbf{The differenced map accounts for the galaxy overdensity, log(1+$\delta_{gal}$), resulting in a median log(1$+\delta_{ssfr/gal}$)} in MB more than four times that in MF.
        }
        \label{fig:Voronoi}
\end{figure*}

In \S \ref{s:observations} we discuss observations, in \S  \ref{s:results} we explain the analysis methods and results, and in \S \ref{s:conclusions} we discuss possible scenarios to explain them. Equivalent widths are in rest-frame units and distances are given in proper units. Magnitudes are in AB. 
We assume a flat $\Lambda$CDM cosmology with $H_{0}=70$ km s$^{-1}$ Mpc$^{-1}$, $\Omega_{M}=0.3$, and $\Omega_{\Lambda}=0.7$.
\section{Observations} \label{s:observations}
\subsection{Target} \label{ss:target}

The supercluster RX J0910 at $z$$\sim$1.10 was first identified in the ROSAT Deep Cluster Survey \citep{1995astro.ph..9056R,2002AJ....123..619S} by the X-ray emission of RX J0910+5422 (hereafter Cluster \tbf{B}). 
A spectroscopic campaign by \cite{2008A&A...489..571T} identified the nearby RX J0910+5419 (hereafter Cluster \tbf{A}), as well as a larger network of filaments. The LSS was subsequently observed as part of the Observations of Redshift Evolution in Large Scale Environments (ORELSE) survey \citep{2009AJ....137.4867L}, an ongoing multi-wavelength campaign studying the environmental dependence of galaxies 
in 18 LSS fields in the redshift range 0.6$\leq$z$\leq$1.3.

Details on previous multi-wavelength observations and reductions for ORELSE are found in \cite{2008ApJ...684..933G}, \cite{Lemaux2012}, and \cite{2013ApJ...763..124R,2017MNRAS.466..496R}.
In this study we incorporate new photometric and spectroscopic observations, which will be described briefly in  \S\ref{ss:photometry} and \S\ref{ss:spectroscopy} and in detail in Tomczak et al.\ (\emph{in prep.}).
With all spectral members (see \S\ref{ss:spectroscopy}) we calculated luminosity weighted centroids, dynamical masses, virial radii, mean redshifts, and velocity dispersions to be [09\fd10\fm44\fs, 54\fd22\fm21\fs] and [09\fd10\fm03\fs, 54\fd18\fm36\fs], 
2.7$\pm2.0\times10^{14}$  $\mathcal{M}_{\odot}$ and 5.0$\pm4.3\times10^{14}$  $\mathcal{M}_{\odot}$, 
0.82$\pm0.2$ $h^{-1}_{70}$ Mpc and 1.01$\pm0.3$ $h^{-1}_{70}$ Mpc, 1.100 and 1.102, 681$\pm$170 km s$^{-1}$ \tbf{(ORELSE galaxies only)} and 840$\pm$244 km s$^{-1}$, for Clusters \tbf{B and A}, respectively. These quantites were calculated using methods described in \cite{Lemaux2012} and \cite{2013ApJ...763..124R,2017MNRAS.466..496R}.
\tbf{X-ray contours (Fig. \ref{fig:Voronoi}) show no indication of dissociated gas in the wake of an initial pass-through indicating that the clusters have not yet collided with each other.}
A dynamical simulation for merging clusters \citep{Dawson2013} adjusted for a pre-merger system according to \cite{2015ApJ...803..108A} indicates that the clusters, whose centers are separated by 4.6 Mpc, have a Time-Till-Collision (TTC) of 6.1$^{+5.6}_{-3.8}$ Gyr (assuming an angle relative to the plane of the sky of $\alpha\leq70$).

\subsection{Imaging and Photometry} \label{ss:photometry} 

Subaru/Suprime-Cam 
imaging of RX J0910 was performed in five optical bands: $B$, $V$, $R_{c}$, $I^{+}$, $Z^{+}$. Near-infrared imaging in the $J$ and $K$ bands from WFCAM/UKIRT, 3.6 \& 4.5$\mu$m from \textit{Spitzer}/IRAC, 
\tbf{and 24$\mu$m from \textit{Spitzer}/MIPS} was additionally taken on the field. 
We perform Spectral Energy Distribution (SED) fitting on observed-frame magnitudes to derive photometric redshifts ($z_{phot}$), stellar masses (log($\mathcal{M}_{*}$ $\mathcal{M}_{\odot}^{-1}$)), specific star formation rates (SSFR$_{SED}$), rest-frame magnitudes ($M_{AB}$), 
V band dust-attenuation (A$_{V}$) \tbf{and rest-frame total IR luminosities (L$_{TIR}$)}. 
Details on reduction, photometry, and SED fitting are found in \cite{2016arXiv160800973L} and Tomczak et al.\ (\emph{in prep.}).
Spectroscopic redshifts were used to determine the $z_{phot}$ scatter (\S\ref{ss:spectroscopy}). Details on this procedure are found in \cite{2016arXiv160800973L}.
The 1$\cdot\sigma_{z}$ cluster $z_{phot}$ range is 1.02$\le$z$_{phot}\le$1.19, which is calculated by subtracting and adding 1$\cdot\sigma_{z}*(1+z_{LSS})$ to the minimum and maximum of the LSS spectroscopic redshift range (\S\ref{ss:spectroscopy}), respectively. 
We find that for Z$^{+}$$\le$23.25, all spectroscopic LSS members have a measured $z_{phot}$ within \tbf{this range.}

\subsection{Spectroscopy}  \label{ss:spectroscopy} 

In total, seven slitmasks were observed through with the \ion{Keck}{II}/DEep Imaging Multi-Object Spectrograph \citep[DEIMOS;][]{2003SPIE.4841.1657F} between 2009-2015, under seeing of 0.50-1.05\arcsec$\ $with an average exposure time of 9704 s per mask. The 1200 l mm$^{-1}$ grating was used with 1\arcsec$\ $wide slits, resulting in a full width half maximum resolution of 1.7 \AA. Central wavelengths were set to 8000-8100 \AA.
These observations resulted in 750 high-quality spectral redshifts \citep[Q=-1, 3, 4;][]{2008ApJ...684..933G,Newman2013}. 

The spectroscopic redshift ($z_{spec}$) range for RX J0910 is determined off the clear peak in the redshift histogram near $z$$=$ 1.10. The redshift range 1.09$\le$$z$$\le$1.12 was 
adopted as it encompasses the \tbf{true} members of both clusters. This redshift range, along with a magnitude cut of Z$^{+}$$\le$23.25 \tbf{($\sim$0.5$L^{\ast}$ for a cluster galaxy at this redshift)}, defines the sample of $z_{spec}$ member
galaxies of RX J0910 and contains 102 galaxies, 100 from ORELSE and two additional members confirmed from \cite{2008A&A...489..571T}.
The above magnitude limit was chosen to optimize completeness based on the number of targeted objects (63\%) brighter than this limit for which we attained high-quality spectra (84\%) within the photometric redshift range (see \S \ref{ss:photometry}).
The area of spectroscopic coverage includes all objects seen as black dots in the left panel of Fig. \ref{fig:Voronoi}. This boundary constrains the outermost gray border in the right panel of Fig. \ref{fig:Voronoi} where it intersects our regions of special 
study defined in \S \ref{s:results}. As a result, only areas internal to the \tbf{boundaries} of spectral coverage contribute to \tbf{the analysis presented in this study.}

\section{Results} \label{s:results}

We define regions in units of virial radii, R$_{vir}$, 
physically motivated by the radial dependence of the Navarro Frenk White \citep{1996ApJ...462..563N} profile and the tidal force this dark matter halo exerts on a galaxy (see discussion in \S \ref{s:conclusions}). R$_{vir}$, as estimated from the velocity dispersion, 
is a reasonable metric due to the long TTC \citep{1996ApJS..104....1P}. 

The member galaxies of clusters A and B, defined by core and infall regions around each centroid, R$_{proj}$$\le$0.5R$_{vir}$ and 0.5 R$_{vir}$$<$R$_{proj}$$\le$1.5 R$_{vir}$, respectively 
are \tbf{excluded from our analysis.} The scale of each cluster is shown by the inner gray circles plotted in Fig. \ref{fig:Voronoi}, where these circles correspond
to R$_{proj}$=1.5R$_{vir}$ for each cluster.
\ewOII, \ewHd and \Dn measurements from composite spectra indicate that the galaxies in the cores of Clusters A and B are undergoing little star formation, which is consistent with the predicted cessation of star formation as galaxies reach the core of massive clusters \cite[e.g.,][]{ButcherOemler78}.
Unlike Cluster \tbf{A}, the infall region of \tbf{B} is active, consistent with the results of \cite{Mei2006}.
The core (9 \& 8) and infall (18 \& 17) members of both clusters (A \& B, respectively) are removed from our analysis to eliminate any confusion from effects related to secular cluster processes. 

The Merger Front (MF) region, outlined in red in Fig. \ref{fig:Voronoi}, is defined as the overlap between circles of R$_{proj}$=4R$_{vir}$ centered on the luminosity-weighted centroid of each cluster 
and thus probes the galaxies most affected by the impending merger. 
The Merger Back (MB) is defined by the areas in the range 1.5R$_{vir}$$<$R$_{proj}\le$4R$_{vir}$ \tbf{centered on} each cluster that do not overlap, although it is contrained by the outer slitmask boundary, and is shown by the outermost gray border in Fig. \ref{fig:Voronoi}. The MF and MB have 17 
and 24 members, respectively, and are the focus of this letter. \tbf{Results in this letter do not change meaningfully if the X-ray centroids are used instead.}

\begin{figure*}
          \begin{center}
          \includegraphics[height=66mm]{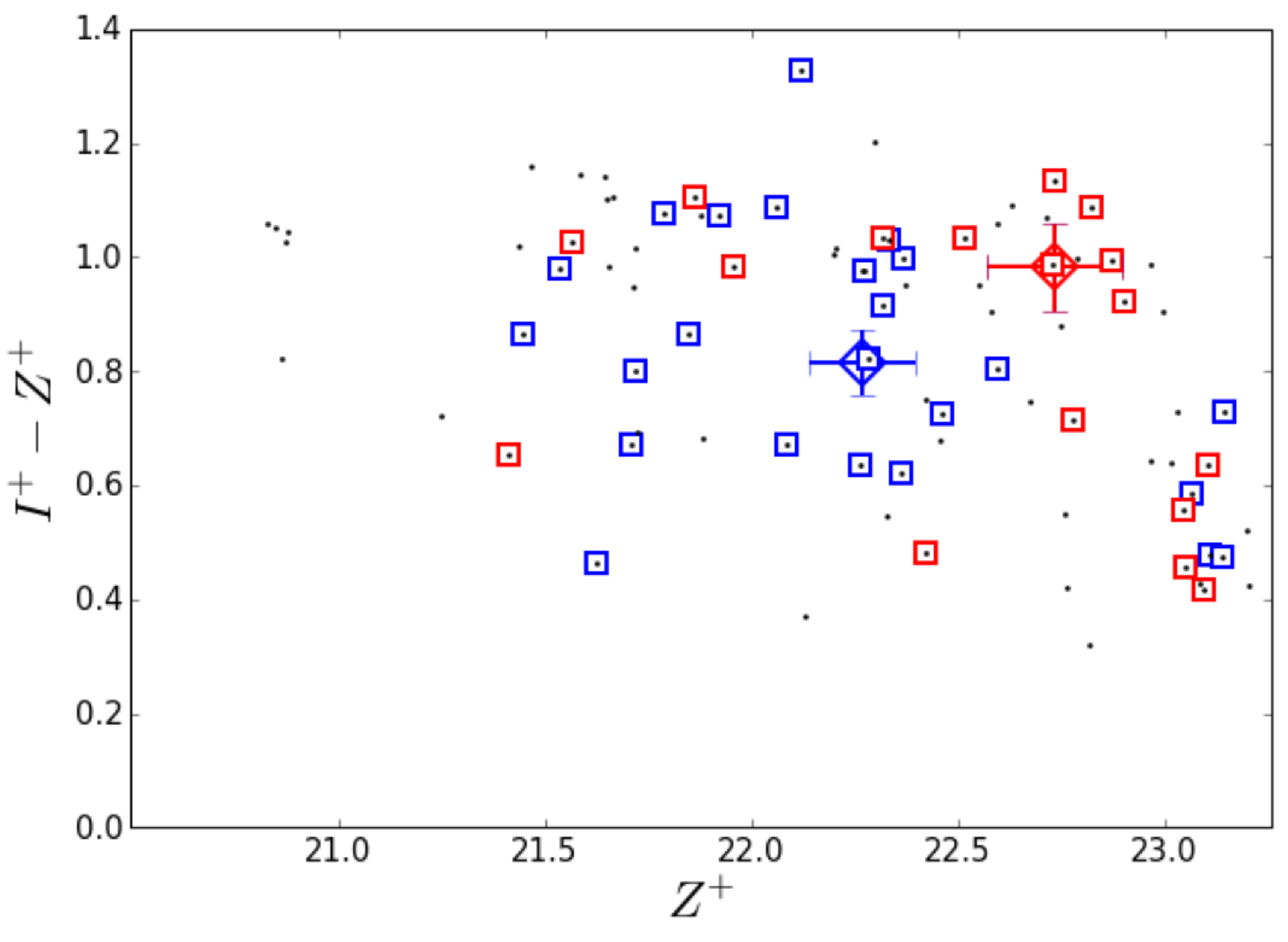}
          \includegraphics[height=66mm]{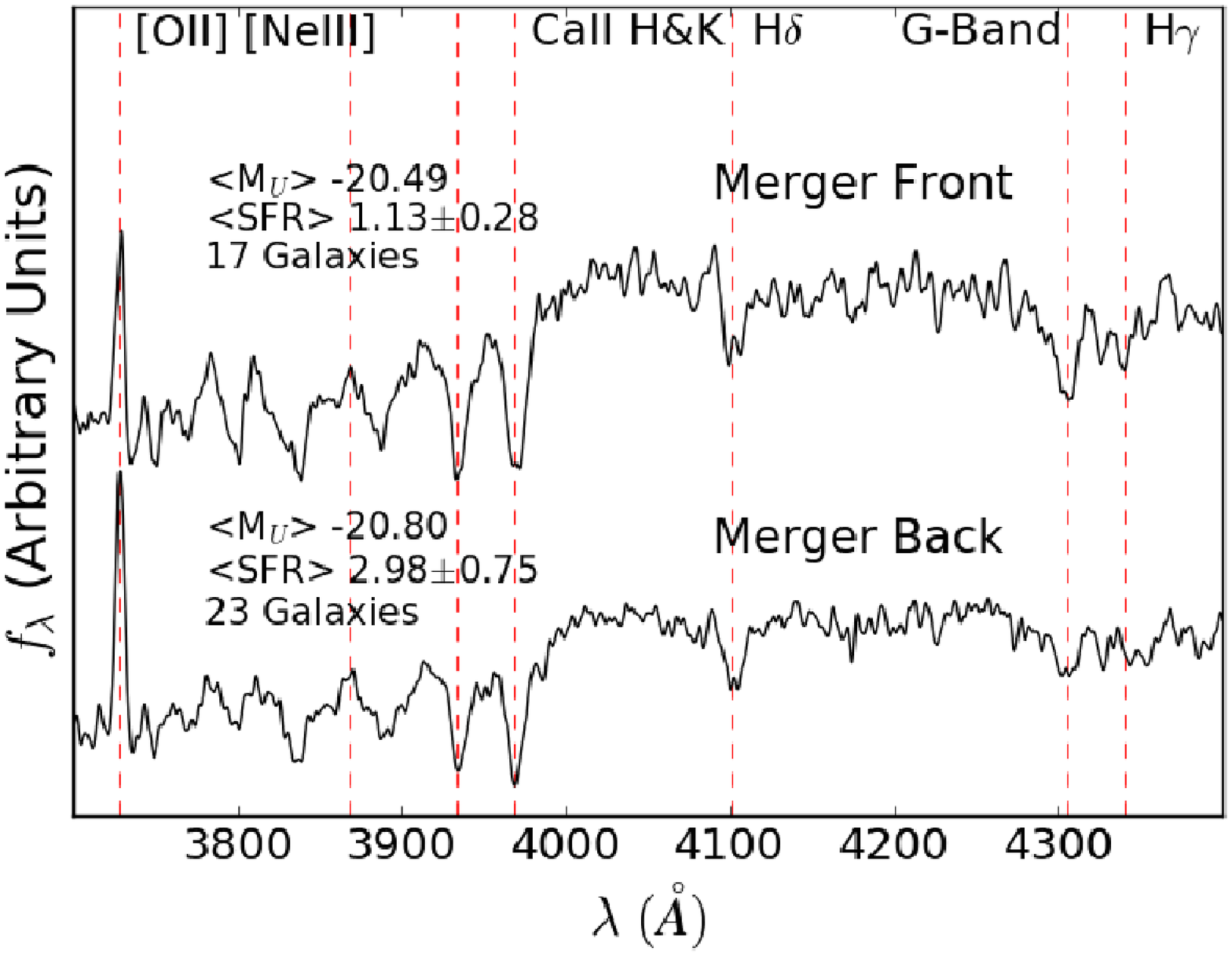}
          \end{center}
        \caption{
        \textit{Left:} Color-Magnitude Diagram of the MF (red squares) and MB (blue squares) galaxies with median and uncertainties overplotted as large symbols (\S \ref{ss:CMD}). Black dots are spectroscopically confirmed members in all regions within the LSS redshift range.
  A KS test \tbf{on the magnitude distributions} reveals that that the galaxy populations of the MF and MB do not originate from the same underlying population.
        \textit{Right:} composite spectra of the 17 MF and 23 MB \tbf{(Type-1 AGN removed,} see \S \ref{ss:spectroscopy}) spectra used to calculate the corresponding extinction-corrected SFR
        SFR (\S \ref{ss:sfr}). Both composites are smoothed with a Gaussian kernel of $\sigma=1.1$ \AA. Average $<$$M_{U}$$>$ and $<$$SFR(L_{[OII]})$$>$ are indicated (Table \ref{tab:1}). Red dotted lines label key spectral features.
        }
        \label{fig:CMD}
\end{figure*}

\subsection{Color-Magnitude Properties} \label{ss:CMD} 

A color-magnitude diagram is a simple yet effective way to examine ensemble properties of a population. 
For this analysis we use the $I^{+}$ and $Z^{+}$ observed-frame magnitudes as they are adjacent filters that capture either side of the 4000\AA$\ $break at $z\sim1.1$ and are not model dependent. 
The difference between the MF and MB distributions is apparent by eye in Fig. \ref{fig:CMD}: the MB population dominates the bright blue region of the CMD traditionally associated with prodigious star formation, whereas MF is primarily composed of red and fainter blue galaxies. 
We confirm that this difference is statistically robust in the following way.
We performed a one-dimensional Kolmogorov-Smirnov (KS) test on the magnitude distributions of MF and MB galaxies finding that they are not drawn from the same underlying distribution at the $>$97\% confidence level. An identical test on the color distributions finds the confidence level at $>$65\%.
Additionally, we find that the median magnitude and color for the MF and MB distributions are disparate at $>$95\% and $>$68\%, respectively. These values and their associated errors are shown in Fig. \ref{fig:CMD}.
Note that the significance of these differences holds if we instead use model-dependent rest-frame magnitudes (i.e., $M_{NUV}-M_{r}$ vs. $M_{r}$). Indeed, further underscoring the differences between the two populations, 
we classify galaxies in MF and MB into quiescent and star forming using the rest-frame $NUVrJ$ separations of \citet{themound14} and find  quiescent fractions ($f_{q}$) of 55\% and 25\%, respectively.

We additionally include $z_{phot}$ members using two methods in order to investigate whether spectroscopic completeness effects could be responsible for the observed differences. 
First we include with the $z_{spec}$ members all objects in the redshift range $\pm$1$\sigma_{z_{phot}}*(1+z_{LSS})$ and $\pm$2$\sigma_{z_{phot}}*(1+z_{LSS})$. 
A KS test reveals that difference holds at $\ge$90\% level and at $\ge$65\% or better confidence for magnitude and color, respectively. 
In our second approach, we integrate the probability distribution function of $z_{phot}$ of each object over the $z_{spec}$ range of the LSS (as in \citealt{2017MNRAS.466..496R}). 
Adopting P{(1.09$<$$z_{phot}$$<$1.12)}$>$0.23 as the criterion for $z_{phot}$ membership as a balance between purity and completeness results in a $\ge$90\% and $\ge$80\% confidence in the difference in the combined $z_{phot}+z_{spec}$ magnitude and color distributions.
Similar tests are run on the stellar mass distributions of the MB and MF galaxies. No such differences exist between the stellar masses of the MB and MF galaxies (with median values of log($\mathcal{M}_{*}$/$\mathcal{M}_{\odot}$)$\sim$10.8 in both cases) 
\tbf{and a KS test shows no significant difference between the distributions}. These results preclude the possibility that any observed SFR regulation is driven by trends with stellar mass \citep[e.g.,][]{2007ApJ...670..156D}.

We conclude that the difference between the color-magnitude distributions for galaxies most and least likely to be affected by the merger is not due to poor completeness, nor due to biased sampling or to stellar mass effects. We proceed under the assumption that 
our spectroscopic data are representative of the true population and continue to suspect that the \tbf{preponderance of red and fainter blue galaxies in the MF may be indicative of a lower average SFR due to a process related to the impending merger.}

\begin{table*}
    \centering
    \caption{Star Formation Metrics}
		\begin{tabular}{p{0.6cm}p{2.7cm}p{1.9cm}p{1.8cm}p{0.4cm}p{0.4cm}p{0.8cm}p{0.4cm}p{0.7cm}p{1.3cm}p{1.5cm}}
        \toprule
 \footnotesize{Region}
 & \footnotesize{Area}
 & \footnotesize{$<$$SFR(L_{[OII]})$$>$}
 & \footnotesize{$\widetilde{(1+\delta_{ssfr/gal})}$}
 & \footnotesize{$\widetilde{Z^{+}}$}
 & \footnotesize{$\sigma^a$}
 & \footnotesize{$\widetilde{I^{+}-Z^{+}}$}
 & \footnotesize{$\sigma^a$}
 & \footnotesize{N$_{spec}^b$}
 & \footnotesize{$<$R$_{tidal}$$>$$^c$} 
 & \footnotesize{SFR$_{[OII]+IR}^d$} \\
 { }
 & {arcmin$^{2}$ (h$^{-1}_{70}$Mpc)$^2$}
 & {$\mathcal{M}_{\odot}\ yr^{-1}$}
 & {}
 & {}
 & {}
 & {}
 & {}
 & {}
 & \footnotesize{kpc} 
 & {$\mathcal{M}_{\odot}\ yr^{-1}$} \\
\midrule
MF & 25.37 (6.34) & 1.13$\pm$0.28 & 0.20 & 22.73 & 0.16 & 0.98 & 0.08 & 17 & 102$\pm$20 & 2.53$\pm$0.29 \\
MB & 97.98 (24.49) & 2.98$\pm$0.75 & 0.83 & 22.27 & 0.13 & 0.82 & 0.06 & 24 & 203$\pm$40 & 5.60$\pm$0.61 \\
\bottomrule
\end{tabular}
\begin{flushleft}
$^a${\footnotesize{Uncertainty on the median for magnitudes and colors (1.253$\cdot\sigma$/$\sqrt(N)$), plotted in Fig. \ref{fig:CMD} with large symbols and error bars (\S \ref{ss:CMD}).}}\\
$^b${\footnotesize{Number of galaxies with a high-quality $z_{spec}$ in the LSS spectroscopic redshift range to a magnitude limit of $Z^{+}$$\le$23.25}}\\
$^c${\footnotesize{\tbf{Tidal radius for test galaxy half the distance between the cluster centroids (the MF) versus the same distance behind each cluster (the MB) using the total and average velocity dispersions, respectively (\S \ref{s:conclusions}). This value depends on the angle of the merger relative to the plane of the sky ($\alpha$) where larger angles bring the ratio of the tidal radii for galaxies in the two regions closer to unity.}}}\\
$^d${\footnotesize{\tbf{SFR$_{[OII]+IR}$ = SFR($L_{[OII], uncorr}$)+SFR$_{IR}$, see \S \ref{ss:sfr}}}}\\
\end{flushleft}
\label{tab:1}
\end{table*}

\subsection{Extinction-Corrected Star Formation Rate} \label{ss:sfr}

Encouraged by the suggestively distinct color-magnitude properties of the MF and MB galaxies, we leverage our sample of high-quality spectra, along with multi-band photometry fitted with stellar population synthesis models, to calculate extinction-corrected SFRs \tbf{using two different methods}.

First, we use the mean values of \ewOII, M$_{U}$, A$_{V}$ and the distance modulus for $z_{spec}$ members of each region to determine L(\OII) and thus an extinction-corrected $SFR(L_{[OII]})$. 
\ewOII can be used as a proxy for star formation when H$\alpha$ is not available \citep{P99}. Though the use of \OII introduces a risk of contamination from non-star forming sources such as Seyferts/Low Ionization Nuclear Emission line Regions (LINERs) \citep{Lemaux2010,2016arXiv160800973L}, we mitigate this risk by removing a type-1 AGN from the sample and by using the additional metrics for star formation presented in this letter that are not subject to the same impurity. 
We make inverse variance, unit-weighted composite spectra for each region, and measure \ewOII using the bandpass method described in \cite{Lemaux2010}. 
$<$M$_{U}$$>$ is used because it provides a fair sampling of the rest-frame continuum surrounding [OII].
\tbf{Extinction corrections are made using $<$A$_{V}$$>$ and adopting the scheme of} \cite{2013ApJ...779..135W} to minimize the scatter and offset between line-measured and SED-fit SFRs, shown to work well with our method \citep{2017A&A...599A..25P}.
The extinction-corrected $<$$SFR(L_{[OII]})$$>$ is \tbf{2.64$\pm0.93\times$ higher for MB than MF, 2.98$\pm$0.75 $\mathcal{M}_{\odot}\ yr^{-1}$ and 1.13$\pm$0.28} $\mathcal{M}_{\odot}\ yr^{-1}$, consistent with the larger fraction of fainter blue galaxies 
within the MF region. This decrease is also consistent with the elevated $f_{q}$ within the MF region, though it is not possible to discern whether the fractional excess of quiescent galaxies is solely responsible for the observed decrease in
the average SFR due to the small number of each galaxy type within each region. 

\tbf{In our second approach, we use L$_{TIR}$ for areas with MIPS coverage to calculate $SFR(L_{TIR})$ (Table \ref{tab:1}) and the surface density of signal-to-noise ratio $>$2 sources, a limit
that corresponds to $\ga$6.5 $\mathcal{M}_{\odot}\ yr^{-1}$ at the redshift of RX J0910.}
\tbf{The results indicate that we are not differentially missing a population of dusty starbursting galaxies in MF.
Further, we re-calculate total SFRs of galaxies in the two regions by combining the extinction-uncorrected $SFR(L_{[OII]})$ with the SFR derived from the median $L_{TIR}$ value for all galaxies in each population. This exercise yields an SFR$_{[OII]+IR}$ that is 
2.21$\pm$0.35 times higher in MB than in MF (Table \ref{tab:1}).}

\subsection{Voronoi Monte Carlo Tessellation} \label{ss:VMC}

The Voronoi Monte Carlo (VMC) technique allows us to utilize both $z_{spec}$ and $z_{phot}$ to define a metric for local environment, on to which we can map galaxy properties like SSFR$_{SED}$ (SFR$_{SED}$/$\mathcal{M}_{\ast}$, \S \ref{ss:photometry}). A Monte Carlo approach to using the Voronoi tessellation method for reconstructing the galaxy density field is developed and extensively tested in \cite{2015ApJ...805..121D}. A similar technique is developed for ORELSE data by \cite{2016arXiv160800973L}. 
We perform 100 MC iterations, each time resampling from a Gaussian constructed from their P($z$) for all galaxies without a secure $z_{spec}$. Confirmed members \tbf{appear in all iterations}. We require that all objects used to generate the maps satisfy 18$<$$Z$$^{+}$$<$24.5, 
SFR$>$10$^{-3}$ $\mathcal{M}_{\odot}\ yr^{-1}$ and 
log($\mathcal{M}_{*}$ $\mathcal{M}_{\odot}^{-1}$)$>$9,
to minimize incompleteness with respect to faint, quiescent, low-mass galaxies.

We calculate the VMC tessellation for both galaxy overdensity, in units of log(1$+\delta_{gal}$), and SSFR$_{SED}$ overdensity, in units of log(1$+\delta_{ssfr}$), within the LSS.
The quantity \tbf{log(1$+\delta_{ssfr}$)} is inherently correlated with galaxy density as \tbf{it is calculated from SSFR$_{SED}$ divided by the Voronoi area,} 
so we decouple the measurements by subtracting the log(1$+\delta_{gal}$) \tbf{map from its log(1$+\delta_{ssfr}$) equivalent} (Fig. \ref{fig:Voronoi}: left) resulting in a differenced map (Fig. \ref{fig:Voronoi}: right). We define the 
\tbf{log(1$+\delta_{ssfr}$)} 
normalized by galaxy density as log(1$+\delta_{ssfr/gal}$), which is \tbf{essentially} a proxy of the overdensity of the SSFR$_{SED}$ per galaxy.
The median (1$+\delta_{ssfr/gal}$) value in the MF is 0.20, more than four times lower than the median in MB (0.83). This result is consistent with our additional star formation indicators from \S \ref{ss:CMD} and \S \ref{ss:sfr}, all suggesting a suppression of star formation in the region between the two merging clusters.

Cluster mergers are frequently housed in filaments, which can introduce quenching mechanisms \citep{2015ApJ...814...84D}. 
\tbf{We examine the case of a filament as the source of suppression in MF by measuring the median (1$+\delta_{ssfr/gal}$) value in those areas
of MB which are aligned with the merger axis, an axis defined by connecting the two cluster centers by a straight line \citep{Dawson2013}. These regions were defined to have 
an equal extent in the dimension perpendicular to the merger axis as the MF region and to be situated on the opposite side of each cluster stretching to the edge 
of th MB region along the dimension oriented with the merger axis.
We find the median log(1$+\delta_{ssfr/gal}$) in the region to be higher relative to MF by roughly the same factor (3.31) as in the original comparison.
This result suggests that filamentary-dependent dynamics within a filament along the merger axis is not the predominant mechanism for suppression in MF.}
\vskip-0.9cm
\section{Discussion} \label{s:conclusions}
Our results, summarized in Table \ref{tab:1}, reveal a consistent series of measurements indicating a dearth of star formation in the region between Clusters A and B relative to the galaxies at the same clusto-centric distance on the leeward side of the impending merger.
Why would a relative dearth of star formation occur in the merger front?


One explanation for this suppression may be the amplified tidal force experienced by galaxies in MF caught between two massive, approaching dark matter potentials.
A galaxy inside a single cluster halo feels a stronger gravitational pull on one side than the other.
The resulting differential force can remove loosely bound gas in the galaxy disk and halo which could otherwise be used for star formation.
\tbf{Using the measured velocity dispersions of the two clusters (\S \ref{ss:target}), we calculate the tidal radius \citep{1998ApJ...495..139M} for a test galaxy, outside which the binding force per unit mass is insufficient to retain material.
Note that this calculation is necessarily conservative as we ignore the tidal effects of sub-halos as well as
those of nearby galaxies. The tidal radius for galaxies in MF is found to be smaller by a factor of two, on average, compared to a galaxy in MB (Table \ref{tab:1}) assuming the merger is transverse to our line of sight. While the MF value exceeds the size of most observed HI disks in the local universe (e.g., \citealt{wang2014}), it is still small enough to
allow for the stripping of the diffuse outer regions of the HI disk and gas associated with larger scale inflows.} 


\tbf{An additional contribution to the suppression of star formation may be a changing cluster potential, effectively creating a tidal impulse in the frame of the galaxy. 
While such an impulse can be created by galaxies moving through the cluster potential, for galaxies in the MF the same conditions are amplified by the convergence of the two merging cluster potentials.} 
As a result, the MF galaxies can experience tidal heating not experienced by the MB galaxies,
estimated as the kinetic energy introduced to the system and resulting in increased velocity dispersion and mass loss. In a changing tidal field, tidal heating can occur at any radius within a galaxy where a peak in the tidal force occurs.
These phenomena powerful enough to alter not only colors (blue to red) but also morphologies (disk to spheroid) \citep{2003ApJ...589..752G,2003ApJ...582..141G,2008MNRAS.387...79V}. 
The varying external tidal force in the MF can result in the stripping of material that would otherwise fuel star formation \citep[e.g.,][]{Larson1980}. 
\cite{2003ApJ...589..752G} simulated the impact of tidal forces and tidal heating on a range of galaxies as they move through various cluster potentials finding effects that can halt star formation in the disks of large spirals and completely destroy low-density galaxies like dwarf spheroidals, sending debris into the surrounding medium. 


Clusters evolve in regions where there is a confluence of dark matter, gas and galaxies, so we must also consider the influence of the greater LSS. 
As discussed in \S \ref{ss:VMC}, 
filaments may have an effect on star formation. Not only does pre-processing kick-start the depletion of star-forming resources before halo accretion, nuclear activity can heat gas, causing it to expand beyond the tidal radius. 
\tbf{While a preliminary test suggests that the filament along the merger axis is not likely the cause of the suppression of star formation in MF, we cannot completely rule out such effects.}

Ultimately, semi-analytical and hydrodynamical simulations are necessary to understand how, in an equal mass merger, disk heating and tidal stripping affect star formation due to the non-linear nature of the galaxy and host halo interactions. 
Such simulations are also necessary to investigate filament dynamics.
In addition, a larger number of pre-merger systems observed at different stages leading up to the first pass-through are necessary to fully assess the plausibility of the scenario proposed in this study.
Such an ensemble would allow for the study of both short- and long-term effects on star formation without sacrificing the knowledge of the initial states of both clusters.
\vskip 0.3cm
\noindent
\textbf{Acknowledgements}

\noindent
{\footnotesize
This material is based upon work supported by the NSF under Grant No. 1411943 and NASA Grant Number NNX15AK92G. This study is also based, in part, on data collected at the Subaru Telescope obtained from SMOKA, which is operated by the ADC at the NOAJ. The spectrographic data presented were obtained at the W.M. Keck Observatory, operated as a scientific partnership among the CalTech, the UC system, and NASA.
We thank the hard-working staff at the facilities used in this letter and the indigenous populations for allowing us to observe on their sacred land. We also thank Nathan Golovich for discussions helpful to this letter.
}





\bibliographystyle{mnras}
\bibliography{MNRAS_paper_amansheim} 


\bsp	
\label{lastpage}
\end{document}